\documentclass[usenatbib]{mn2e}
\usepackage{amssymb}
\usepackage{amsmath}
\usepackage{ctable}
\usepackage{fixltx2e} 
\usepackage[utf8]{inputenc}
\usepackage{hyperref}
\usepackage{csquotes}
\usepackage{xspace} 
\usepackage{listings}
\usepackage{float}

\def\avisc{{\alpha_{\rm visc}}}
\newcommand{\degree}{^{\circ}}

\def\msun{{\rm M_{\odot}}}
\def\mbh{M_{\rm BH}}
\def\mshell{M_{\rm shell}}
\def\mdisc{M_{\rm disc}}
\def\mcore{M_{\rm core}}
\def\rcore{r_{\rm core}}
\def\racc{r_{\rm acc}}
\def\rcirc{r_{\rm circ}}
\def\rin{r_{\rm in}}
\def\rout{r_{\rm out}}

\def\tilt{\theta_{\rm tilt}}

\def\vrot{{v}_{\rm rot}}
\def\vturb{{v}_{\rm turb}}

\def\avisc{{\alpha_{\rm visc}}}

\def\kms{\rm \, km \, s^{-1}}
\def\myr{\rm \,Myr}

\def\msunpc3{\,\msun~{\rm {pc^{-3}}}}
\def\msunpc2{\,\msun~{\rm {pc^{-2}}}}
\def\msunyrpc2{\,\msun~{\rm {pc^{-2}}}~{\rm yr^{-1}}}

\def\kms{{\rm\,km\,s^{-1}}}
\def\pc{\mbox{\, pc}}
\def\kpc{\mbox{\, kpc}}

\newcommand{\pgadget}{\textsc{P-Gadget}\xspace}

\def\simlt{\mathrel{\rlap{\lower 3pt\hbox{$\sim$}}\raise 2.0pt\hbox{$<$}}}
\def\simgt{\mathrel{\rlap{\lower 3pt\hbox{$\sim$}} \raise 2.0pt\hbox{$>$}}}
\def\lsim{\mathrel{\rlap{\lower 3pt\hbox{$\sim$}}\raise 2.0pt\hbox{$<$}}}
\def\gsim{\mathrel{\rlap{\lower 3pt\hbox{$\sim$}} \raise 2.0pt\hbox{$>$}}}

\newcommand{\acknowledgments}{\begin{small}\section*{Acknowledgments}\end{small}}
\newcommand\altaffilmark[1]{$^{#1}$}
\newcommand\altaffiltext[1]{$^{#1}$}
\title[AGN fuelling by Overlapping Inflows]{
Overlapping inflows as catalysts of AGN activity - II: Relative
importance of turbulence and inflow-disc interaction
}

\author[Carmona-Loaiza et al.]{
\parbox[t]{\textwidth}{ 
\vspace{-1.0cm}Juan M. Carmona-Loaiza\thanks{E-mail:jcarmona@sissa.it}\altaffilmark{1}
Monica Colpi\altaffilmark{2,3},
Massimo Dotti\altaffilmark{2,3}
\&\ Riccardo Valdarnini\altaffilmark{1,4} 
} 
\vspace*{6pt} \\
\altaffiltext{1}{Scuola Internazionale Superiore di Studi Avanzati, 
  Via Bonomea 265, 34136 Trieste, Italy.} \\
\altaffiltext{2}{Dipartimento di Fisica G. Occhialini, Universit\`a
  degli Studi di Milano Bicocca, Piazza della Scienza 3, 20126 Milano,
  Italy.} \\
\altaffiltext{3}{INFN, Sezione di Milano-Bicocca, Piazza della Scienza 3, 20126 Milano, Italy. }\\
\altaffiltext{4}{INFN, Trieste - Iniziativa Specifica QGSKY, Italy. } \\
\vspace{-0.7cm} 
}

\begin{document}
\maketitle
\label{firstpage}

\begin{abstract}
The main challenge for understanding the fuelling of supermassive black holes in active galactic nuclei is not to account for the source of fuel, but rather to explain its delivery from the boundaries of the black hole sphere of influence (10-100 pc) down to sub-parsec scales. In this work, we report on a series of numerical experiments aimed at exploring in further depth our model of ``overlapping inflow events'' as catalysts for rapid accretion, seeding a turbulent field in the infalling gas. We initially set a gaseous shell in non-equilibrium rotation around a supermassive black hole. After infall, the shell stalls in a disc-like structure. A second shell is then set in either co-rotation or counter-rotation with respect to the first and is let to impinge on the previously-formed disc. We find that combined turbulence and overlap significantly enhance accretion in counter-rotating inflows, while turbulence dominates for co-rotating inflows. The leftovers of overlapping inflows are warped nuclear discs, whose morphology depends on the relative orientation and angular momentum of the disc and the shell. Overlapping inflows leave observational signatures in the gas rotation curves.
\end{abstract}
\vspace{-5.0 cm}

\section{Introduction}

Understanding accretion onto super-massive black holes (SMBHs), from galactic scales of several kpc down to the horizon scale of few $\mu$-pc, is critical for understanding their growth and determining their relation with the host galaxy (\citealt{Soltan82, Salucci99, King03, Shankar04, Marconi04, Merloni04, Hopkins06eat, Raimundo09}). Gravitational instabilities, either internal or triggered by mergers, are believed to be the main catalysts of SMBH growth over cosmic time, giving rise to the so called SMBH-galaxy co-evolution scenario (\citealt{Ferrarese00,Gebhardt00, Ferrarese05,Kormendy13}).

Once gas makes its way to the sub-parsec region, magneto-hydrodynamical processes, as discussed by \cite{Balbus98}, drive gas down to the innermost stable orbit of the SMBH. However, AGN accretion discs must be continuously or episodically replenished on these scales, because at larger distances the discs are expected to fragment due to self-gravity  (e.g. \citealt{Shlosman90, Goodman03,Lodato12}).  For thin accretion discs, with aspect ratio $H/R \simeq 10^{-3}$, gravitational instabilities start being important at a distance of the order of $R_{\rm SG} \simeq 10^{3} R_{\rm S}$, where $R_S$ is the Schwarzschild radius. For a black hole of $M_{\rm BH} = 10^8 M_{\odot}$, $R_{\rm SG}$ is equivalent to 0.01 pc. However, it is still a matter of debate to find suitable mechanisms that can make the gas cross the angular momentum barrier in the vicinity of the  gravitational influence radius of the BH, $R_{\rm infl} = G \mbh / \sigma^2 \simeq 11 (\mbh / 10^8 \msun) \,\rm pc$, where $\sigma\sim 100$ km s$^{-1}$ is the velocity  dispersion of the stellar component. 

Several mechanisms able to bridge this gap have been proposed in the literature (see \citealt{Alexander12}), which basically fall into two groups (\citealt{Bellovary13}): either i) the gas falls almost radially in cold filaments from large distances leaving a sizeable amount of gas close to the SMBH (\citealt{Bournaud11, DiMatteo12, Barai12, Gaspari13, Feng14}) or ii) the gas loses angular momentum by a series of processes such as gravitational instabilities, stellar feedback, turbulence, disordered flows or the formation of nested bars (\citealt{Wada04, Jogee06, Lodato06, Kawakatu08, Hopkins10sim, Choi13}). In particular and interestingly for the purposes of this paper, \cite{Hobbs11} have shown that supersonic turbulence is a key feature for enhancing the fuelling rate onto a SMBH from 100 pc scales down to the central parsec, because it creates convergent high density flows. Dense filaments are set into ballistic trajectories that travel through the ambient gas, ultimately hitting the inner boundary.  \cite{Hobbs11} model their flows as initially unstable rotating spherical shells to capture the physics of this basic process. 

In a previous study (\citealt{Carmona-Loaiza14}; Paper-I hereafter), we explored the hydrodynamics of a converging non-turbulent spherical shell of gas impacting on a pre-existing stalled gaseous disc, adopting the initial condition proposed by \cite{Hobbs11}. This model, albeit highly idealised,  was constructed to study how the interaction and collision of a converging flow onto a disc associated with a previous ``failed'' episode of accretion would enhance the accretion rate onto a SMBH, to complement the \cite{Hobbs11} model. The disc was not implanted in situ and constructed from an equilibrium model, but generated from an infalling shell with a given angular momentum.  The level of co- or counter-rotation  between the impacting fluid and the disc was, not surprisingly, found to play a key role in determining the amount of gas dragged down to sub-pc scales: a major inflow of gas is triggered when the shell is counter-rotating relative to the disc. The interesting aspect is that not all of the gas is dragged inwards, but a remnant of nested rings, warped and inclined at various angles, is left.  In contrast with the usually proposed mechanisms, the overlapping inflow scenario provides a mechanism not of angular momentum redistribution through torques but rather of angular momentum cancellation via shocks.

In this paper, we aim at studying the role of turbulence in misaligned overlapping inflows, and at highlighting potential observational signatures of the fuelling mechanism proposed. We test the stability of the stalled nuclear disc against self-gravity in the Appendix \ref{sec: SG}, as its fragmentation would reduce the hydrodynamical interaction between the flows. These tests allow us to select the maximum mass for which fragmentation is unimportant within our simple model. Our suite of hydrodynamical simulations is constructed considering as main parameters;  i) the initial angular momentum of the uniformly rotating shell, proportional to its rotation velocity $\vrot$;  ii) the relative orientation between the angular momenta of the disc and the shell, $\theta_{\rm tilt}$; iii) the level of turbulence seeded in the fluid, measured by $v_{\rm turb} = \{ 0,\vrot \}$; and  iv) the self-gravity of the disc, related to the ratio $m = M_{\rm disc}/\mbh$. Actual astrophysical inflows are  far more complex (see e.g. \citealt{Wada04}), however, to pursue such a task, one must be able to discern how gas behaves under changes of the most general parameters studied here. Turbulence plays the critical role of widening the angular momentum distribution of the gas in both the disc and the shell, enlarging the cross section of interaction between them, and the outcome of this interaction cannot be predicted in advance.

The paper is organized as follows: In section \ref{sec: IC} we describe the model and the computational method. In particular we illustrate the implementation of the turbulent velocity spectrum and of the time dependent artificial viscosity in the code.  In section \ref{sec: results} we collect the results of all the simulations to quantify the amount of mass crossing our inner boundary around the SMBH, and in section \ref{sec: relics} we describe the flow dynamics of the turbulent shells in the presence of stalled nuclear discs. We focus on the morphologies obtained and how they relate to the angular momentum content of the shell relative to the disc. A way of observing these relic structures through the analysis of the line of sight velocity is discussed in section \ref{sec: OS} and, finally, we draw the conclusions of the study in section \ref{sec: concl}. Appendix \ref{sec: SG}, is devoted to testing the impact of self-gravity on the sub-pc inflow rate and on disc formation from single infalling shells.

\section{Initial Conditions}
\label{sec: IC}

The environment in which our system evolves resembles that of the nuclear region of a galactic bulge. Specifically, a SMBH is located at the origin of an isothermal non-singular spherical stellar bulge described by a mass distribution 
\begin{equation}\label{eq: potential}
M(r) = \mbh + 
 \begin{cases}
   \mcore (r / \rcore)^3, &r < \rcore \\
    M_{\rm bulge} (r/r_{\rm bulge}),  &r \geq \rcore,
  \end{cases}
\end{equation}

\noindent with  $\mbh = 10^8 \msun$, $\mcore = 2 \times 10^{8} \msun$, $M_{\rm bulge} = 10^{10} \msun$, $\rcore = 20 \pc$, $r_{\rm bulge} = 1 \,\kpc$.  This mass distribution does not represent an active component in our simulations. It is just a static background potential into which the gas is immersed to explore the gas dynamics. While it has been shown that a stellar component can actually exert gravitational torques on the gas (\citealt{Hopkins10sim, Chang07}), we here focus on a particular mechanism, not of angular momentum transport but instead of angular momentum cancellation. The addition of a live potential (i.e. stellar particles) would introduce other angular momentum transfer processes which would make less clear the impact caused only by the interaction between the inflowing shell and the nuclear disc.

 We model the gas as a set of $N_{\rm SPH} \simeq 10^6$ SPH (Smoothed Particle Hydrodynamics) particles, in which half of them belong to the nuclear disc and the other half belong to a shell surrounding it which is already unstable to inflow. For the simulations in which just a single shell is evolved, $N_{\rm SPH} \simeq 5\times 10^5$. We introduce an accretion radius $\racc = 1 \pc$  surrounding the SMBH. Gas particles crossing this radius with low enough mechanical energy and angular momentum are regarded as eventually accreted by the SMBH within a few orbital times and thus are removed from the simulation (\citealt{Bate95}).

The initial conditions for the overlapping inflows are very similar to those presented by \cite{Carmona-Loaiza14}, this time giving the shell an additional turbulent velocity field superimposed on the rotational one in the way described by \cite{Hobbs11}. The initial conditions of each simulation consist of a disc-like structure which extends from the accretion radius $\racc = 1 \pc$ out to $r \simeq 20 \pc$, and a  spherical shell with uniform density which extends from $\rin = 30 \pc$ to $\rout = 100 \pc$. Each shell is given a cylindrical velocity profile $v_{\phi} = \vrot = {\rm const.}$, which can take the values $\vrot \simeq 42, 63$ or $146 \kms$ for the different cases treated here. In addition to the rotational velocity, the particles forming the shell are given a turbulent kick spectrum. The mean turbulent velocity is not a free parameter and is tied to the rotation velocity: $\vturb = \vrot$. The turbulent velocity chosen in this way is such that it maximises the angular momentum distribution without overtaking the dynamics (see \citealt{Hobbs11}).

The disc is always generated from a shell like the one just described having initially $\vrot = \vturb = 63 \kms$. This single shell  (i.e. without having a disc interior to it) evolves as it is not in equilibrium initially.Gas falls towards the centre and angular momentum mixing due to shocks combined with the turbulent velocity field, leads to the formation of an extended disc (as opposed to the narrow ring in \citealt{Carmona-Loaiza14}). After a time of $t \simeq 1.5 \myr$ the disc reaches a stationary angular momentum distribution and the accretion rate becomes almost constant in time. At this time,  we stop the simulation and surround the disc by a new uniform-density shell like the one from which the disc was born. The new shell, apart from having a different rotation and turbulent velocity, has its rotation axis tilted with respect to the axis of rotation of the disc. The latter defines the $z$ axis in all of our simulations, and so the disc initially lies in the $x-y$ plane. We will refer to the disc in its initial condition as the {\it primitive disc} throughout this work, and its particles will be distinguished from shell particles.

An isothermal equation of state has been assumed, $P = c_s^2 \rho$, with a constant temperature of $T = 10^3 {\rm K}$. To focus on the hydrodynamical interaction between the two streams of gas under study, we regard the gas as non-self gravitating. Nevertheless we perform six additional simulations of single shells taking into account the self-gravity of the gas to asses its effects and the validity of the non-self-gravity approximation. In the shell + disc simulations the mass of the shell is equal to  the mass of the disc, $M_{\rm shell} = M_{\rm disc}$. The units of mass, length and time that we will use for our computations are $M_u = 10^{10} \msun$, $L_u = 1 \kpc$, and $T_{u} \simeq 5 \, \myr$ respectively. The corresponding unit of velocity is $V_{u} \simeq 208 \kms$ and the gravitational constant is automatically set $G = 1$. From now on we will continue to use these code units unless otherwise specified.

The turbulent velocity spectrum was given as a Gaussian random field with a Kolmogorov power spectrum  \citep{Kolmogorov41}, 

\begin{equation}
P_{v}(k) \sim k^{-11/3},
\end{equation}

\noindent where $k$ is the wave number, since this is still the simplest description that we have of a turbulent fluid (see however \cite{Federrath13} and references therein). The key assumption here is that the velocity field is homogeneous and solenoidal, so that we can define $\vec{v}$ in terms of a vector potential $\vec{A}$ such that $\vec{v} =\nabla \times \vec{A}$. As such, $\vec{A}$ is also described by a Gaussian random field with an associated power spectrum,

\begin{equation}
P_{A}(k) \sim k^{-17/3}.
\end{equation}

\noindent To avoid a divergence in the variance of $|A|$ at small $k$, we introduce a cut-off $k_{\min}$, so that

\begin{equation}
P_{A}(k) = C( k^2 + k^{2}_{\min})^{-17/6},
\end{equation}

\noindent where $C$ is a normalization constant given to ensure that, given a box of volume $8\,r_{\rm shell}^3$ containing the shell,

\begin{equation}
  \frac{8 \pi \, r_{\rm shell}^3}{(2\,\pi)^3} \int_{k_{\rm min}}^{k_{\rm max}} P_{A}(k) \, k^2\, {\rm d}k = \frac{3}{2} \vturb.
\end{equation}

\noindent Physically the cut-off $k_{\min}$ can be interpreted as the scale $\lambda_{\max} = 2 \pi / k_{\min} \sim r_{\rm shell}$, the largest scale on which the turbulence is likely to be driven and $k_{\rm max} = 10^4 k_{\rm min}$ was introduced to cut the integral when convergence has been reached. 

To generate the statistical realisation of the turbulent velocity field we first sample the vector potential $\vec{A}$ in Fourier space using a periodic cubic grid of dimension $256^3$. We draw the amplitudes of the components of $\vec{A}_k$ at each point $(k_x, k_y,k_z)$ from a Rayleigh distribution with a variance given by $|A_k|^2$ and assign phase angles that are uniformly distributed between $0$ and $2 \pi$. We then Fourier-transform the vector potential to real space and take its curl to obtain the components of the velocity field. Finally, for estimating the turbulent velocity field at the position of each SPH particle, we use tricubic interpolation. The parameters of each simulation are given in Table \ref{tab: parameters}.

It is worth saying that the precise form of the power spectrum is irrelevant for the scenario studied, as the importance of turbulence lies in widening the final distribution of the gas. Nevertheless, the power spectra of supersonic solenoidal turbulence and subsonic turbulence have the same scaling relation with one important difference: while in subsonic turbulence the energy spectrum of the velocity field $P[\vec{v}] \propto k^{-5/3}$, in supersonic (solenoidal) turbulence, it is not the velocity field that has the same scaling, but the density-weighted velocity field, such that $P[\rho^{1/3} \vec{v}] \propto k^{-5/3}$ (\citealt{Lighthill55, Henriksen91, Fleck96, Kritsuk07,Federrath13}). Since in our initial conditions the gaseous shell is homogeneous and the velocity field is solenoidal, the energy spectrum to use is indeed indistinguishable from that predicted by Kolmogorov for incompressible homogeneous turbulence.

\begin{table}
\caption{Parameters of the simulations: 4 simulations of single shells with the same specific angular momentum content (given by $\vrot$) and varying mass ($\mshell$) made in order to asses the impact of self-gravity; 6 simulations of single shells of the same $\mshell$ with varying $\vrot$ and turbulent velocity ($\vturb$), to quantify the relevance of turbulence for dragging mass towards the central parsec; 12 more simulations of overlapping inflow events in near co- and counter-rotation ($\tilt = 60\degree$ and $150\degree$ respectively) with and without a turbulent velocity field to isolate the boost to the inflow rates produced by the overlapping of the inflows.}
\centering
\begin{tabular}{|c|c|c|c|c|c|}\hline
run & $ \vrot $ & $\vturb$ & $\tilt$ & $M_{\rm shell} / \mbh$\\ 
\hline
\hline
v03-M2 & 0.3 & 0.3  &  -- &$0.01$\\
v03-5M2 & 0.3 & 0.3  &  -- &$0.05$\\
v03-M1 & 0.3 & 0.3  &  -- &$0.1$\\
v03-M0 & 0.3 & 0.3  &  -- &$1.0$\\
             &       &                        &        &       \\
v02-nt & 0.2 & --  &  -- & $0.1$\\
v03-nt & 0.3 & --  &  -- & $0.1$\\
v07-nt & 0.7 & --  &  -- & $0.1$\\
             &       &                        &        &       \\
v02 & 0.2 & 0.2  &  -- & $0.1$\\
v03 & 0.3 & 0.3  &  -- &$0.1$\\
v07 & 0.7 & 0.7  &  -- &$0.1$\\
              &       &                        &        &       \\
v02t060-nt & 0.2 & --  &  $60\degree$ & $0.025$\\
v02t150-nt & 0.2 & --  &  $150\degree$ & $0.025$\\
v03t060-nt & 0.3 & --  &  $60\degree$ & $0.025$\\
v03t150-nt & 0.3 & --  &  $150\degree$ & $0.025$\\
v07t060-nt & 0.7 & --  &  $60\degree$ & $0.025$\\
v07t150-nt & 0.7 & --  &  $150\degree$ & $0.025$\\
                &       &                        &        &       \\
v02t060 & 0.2 & 0.2  &  $60\degree$ & $0.025$\\
v02t150 & 0.2 & 0.2 &  $150\degree$ & $0.025$\\
v03t060 & 0.3 & 0.3  &  $60\degree$ & $0.025$\\
v03t150 & 0.3 & 0.3 &  $150\degree$ & $0.025$\\
v07t060 & 0.7 & 0.7  &  $60\degree$ & $0.025$\\
v07t150 & 0.7 & 0.7 &  $150\degree$ & $0.025$\\
\hline
\end{tabular}
\begin{flushleft}
\end{flushleft}
\label{tab: parameters}
\end{table}


We used the Lagrangian SPH code \pgadget (\citealt{Springel05gadget, Dolag05}) with a time dependent artificial viscosity (AV) as described in \cite{Morris97} (see also \citealt{Valdarnini11}). In its standard formulation, the artificial viscosity $\avisc$ always acts when modelling turbulent flows, as a good deal of convergent streams exist. However, when applied to turbulent (supersonic) fluids, the expectation is that the standard viscosity ends up damping the turbulent velocity spectrum unphysically fast. Thus, we include a time dependent artificial viscosity $\avisc(t)$: a viscosity parameter that rises to non-negligible values only where and when it is needed (i.e. at a shock), decreasing to almost zero anywhere else. In this scheme, each particle is assigned its own AV parameter, $\alpha_i$ which is evolved according to the differential equation:

\begin{equation}
\frac{{\rm d}{\alpha}_i}{{\rm d}t} = - \frac{\alpha_i - \alpha_{\rm min}}{\tau_i} + S_i.
\end{equation}

\noindent Thus, $\alpha_i$ always decays to $\alpha_{\rm min}$ with $e$-folding time $\tau_i$, while the source term $S_i$ will make its value rise rapidly when approaching a shock. For our isothermal equation of state $P = c_{s}^{2} \rho$,  the decay time-scale is given by (\citealt{Price10}):

\begin{equation}
\tau_i = \frac{h_i}{\sigma c_i},
\end{equation}

\noindent where $\sigma$ is a non-dimensional free parameter which determines over how many information-crossing times the viscosity decays. For all of our simulations we have taken $\sigma = 0.1$.

\noindent The source term is implemented as

\begin{equation}
 S_i = S_0 \, f_i \max(0,-\nabla \cdot \vec{v}_i),
\end{equation} 

\noindent where $S_0$ is another free parameter that we set to $S_0 = 1.0$. A maximum value for the viscosity parameter is also introduced to keep the viscosity bounded between $\alpha_{\rm min} = 0.02$ and $\alpha_{\rm max} = 1.0$.

The (hydrodynamical) smoothing, $h$, and (gravitational) softening, $\varepsilon$, are both adaptive and coupled in such a way that $\varepsilon = h$. We impose a minimum softening dependent on the mass of the shell simulated (in all cases $\varepsilon_{\rm min} \leq 0.1 \pc$). In Appendix \ref{sec: SG} we select the maximum mass for which fragmentation of the disc is unimportant within our simple model. For the rest of the simulations in which calculations were done neglecting self-gravity the gravitational softening plays no role and the smoothing length is fully adaptive with no threshold $h_{\rm min}$ being imposed.

Table~\ref{tab: parameters} lists the three sets of simulations,  the first for a single converging shell, the second describing the interaction of the shell with a pre-existing disc with no turbulence, and the third with turbulence included. The results of the first set with self-gravity are illustrated in the Appendix and motivate the choice of using as reference value for the mass of the shell and the disc, $M_{\rm shell}=\mdisc=0.025 M_{\rm BH}$.

\section{Turbulent versus  impact inflow mode}
\label{sec: results}

\begin{figure}
\includegraphics[width=0.49\textwidth]{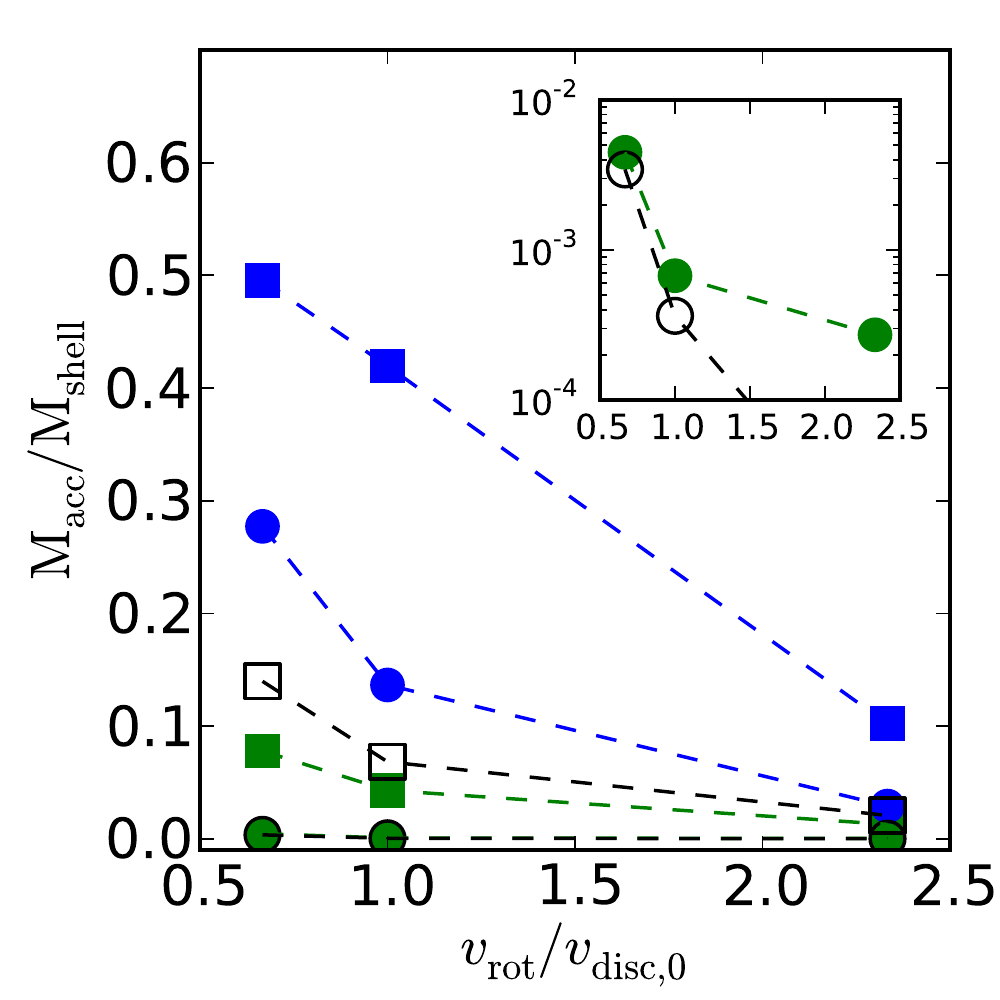}
\caption{Fraction of the mass (in units of $\mshell$) in the central parsec as a function of the initial rotational velocity for all of the simulations in this work and those of Paper-I. Colours blue and green refer to shells falling in co- and counter-rotation with the disc; circles and squares refer to the absence or presence of a turbulent kick in the initial conditions; empty symbols refer to shells falling in the absence of a nuclear disc. Thus, from bottom to top, the data points joined by lines represent the following: i) single shells with no turbulence (\textbf{empty circles}; $<10^{-2}$; shown also in the inset), ii)  co-rotating overlapping inflows with no turbulence (\textbf{green circles}; $< 10^{-2}$; shown also in the inset), iii) co-rotating overlapping turbulent inflows (\textbf{green squares}), iv) turbulent single shells (\textbf{empty squares}), v) counter-rotating overlapping inflows with no turbulence (\textbf{blue cirlces}), vi) counter-rotating overlapping turbulent inflows (\textbf{blue squares}). It is evident that the presence of a nuclear disc  significantly enhances the mass flowing into the central parsec when the $\sim 10 -100 \pc$ scale inflow is counter-rotating with it. See text for details.}
\label{fig: compare}
\end{figure}


Figure \ref{fig: compare} summarises our main result using as key indicator the ``accreted'' mass $M_ {\rm acc}$. We plot the fraction of $M_{\rm shell}$ that crossed the central parsec after $t = 0.3$, versus the initial rotation velocity of the shell, $v_{\rm rot}$ (expressed in units of the rotational velocity of the shell which formed the stalled disc, $v_{\rm disc,0}=0.3$) for the entire data set (See table \ref{tab: parameters}). From the figure we can infer two different modes of angular momentum cancellation and redistribution: One driven mainly by  turbulence --referred to as  turbulent mode, and one driven by the impact of the shell with the disc --referred to as impact mode.

\noindent The data set comprises cases in which: 

\noindent
$\bullet$ (I) a \textbf{single non-turbulent} shell falls inwards onto the SMBH (black empty circles); 

\noindent
$\bullet$ (II) a \textbf{single turbulent} shell falls inwards onto the SMBH as in \cite{Hobbs11} (black, empty  squares);

\noindent
$\bullet$ (III.a) a \textbf{non-turbulent} shell interacts with the stalled disc in near \textbf{co-rotation} (green, filled circles);

\noindent
$\bullet$ (III.b) a \textbf{turbulent} shell interacts with the stalled disc in near \textbf{co-rotation} (green, filled squares); 

\noindent
$\bullet$ (IV.a) a \textbf{non-turbulent} shell interacts with the stalled disc in near \textbf{counter-rotation} (blue, filled circles); 

\noindent
$\bullet$ (IV.b) a \textbf{turbulent} shell interacts with the stalled disc in near \textbf{counter-rotation} (blu, filled squares). 

The sets of parameters that deliver the smallest fraction of gas towards the central parsec are those with no turbulence and no interaction (I). Right above these --although not noticeable in the Figure, non turbulent shells in co-rotation with a stalled disc (III.a) deliver just $\lsim 10^{-2} \mshell$ of gas to the central parsec. The yield is increased to as much as $15\%$ of $\mshell$ for turbulent single shells (II). The interaction with the stalled disc in near co-rotation lowers the accreted mass of the turbulent shells (cases III.b). This was not expected and can be explained
noting that  the denser, higher angular momentum fluid elements present in the disc drag and accelerate the more rarefied fluid elements. These shell elements with lower angular momentum are those that happen to interact soonest with the disc, and this  prevents them from reaching the accretion radius. 

The two most efficient sets of parameters for driving gas to the central parsec are those in which the inflow is driven by the impact-mode (cases IV.a and IV.b). In the cases (IV.a), up to $1/3$ of the shell's  gas is poured into the central parsec, and the yield is raised up to $M_{\rm accc}\sim 1/2 \mshell$ in the cases (IV.b). Even for the shells with the highest angular momentum ($\vrot/v_{\rm disc,0} = 2.33$), the impact-mode of accretion drags a non-negligible $M_{\rm acc}\sim 0.1 \mshell$ amount of gas into the central parsec. Even if their high angular momentum would intrinsically make these shells deliver a negligible amount of mass into the central parsec, the interaction with the disc raises this number to levels comparable with those produced by lower angular momentum shells. 

These considerations hold true if  the masses involved are low enough for the nuclear discs not to disrupt (see Appendix \ref{sec: SG}), as more massive clumpy discs would be expected to evolve into stellar discs and consequently would not exert the required hydrodynamical torques for triggering sub-pc inflows from the $100 \pc$ scales. It is the presence of the \textit{gaseous} disc  that causes shell fluid elements impacting on it to cancel their angular momentum, ultimately driving  gas towards the central parsec. Turbulence magnifies this effect by i) creating converging flows, as described by \cite{Hobbs11}, and by ii) widening the section of interaction between the shell and the disc.

\section{Relic Structures}
\label{sec: relics}

In this section, we describe the properties of the relic structures that form after the violent shell-disc interaction.

The key role of turbulence is that of broadening the disc formed by the first infalling shell. This is because turbulence widens the initial angular momentum distribution (\citealt{Hobbs11}), causing gas to circularise in lower as well as in higher angular momentum orbits relative to the mean circularization radius. Some of this gas crosses the accretion radius and some remains at the periphery forming a wide disc. 
When a second shell interacts with the relic disc, the collision proceeds roughly as described in Paper-I: the shell impacts the disc and the angular momentum of both disc and shell change under the condition of angular momentum conservation of the shell+disc system. This time however, the interaction takes place over a wider radial range.
Thus, while the structures studied in Paper-I lead to the formation of tilted nested rings, in this work the presence of an initial turbulent kick leads to the formation of continuous warped discs surrounded by a more extended and dynamically decoupled disc. As we have seen, the inflow rates are also enhanced. 

\subsection{Counter-rotating flows}

\begin{figure*}
\begin{tabular}{cc}
\includegraphics[width=0.3\textwidth]{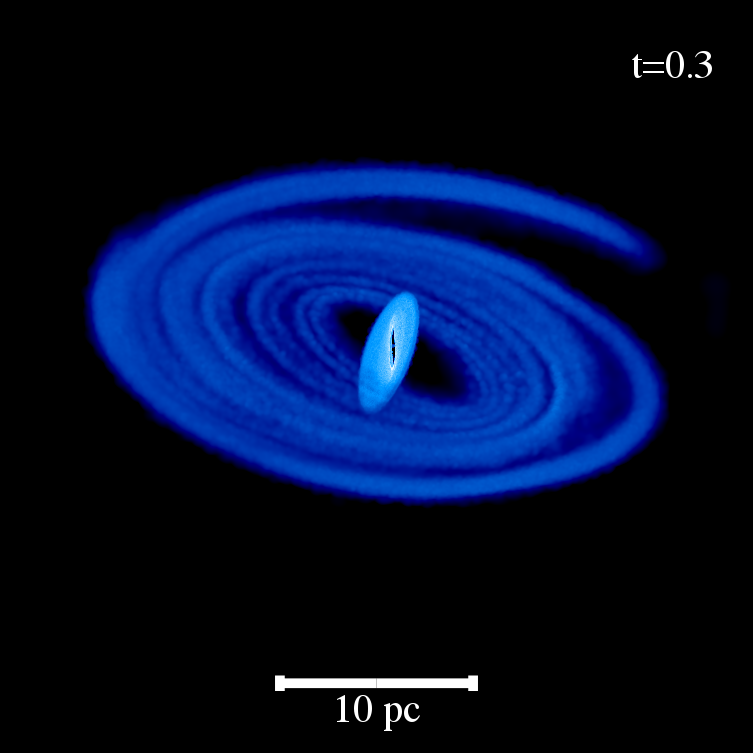}
\includegraphics[width=0.3\textwidth]{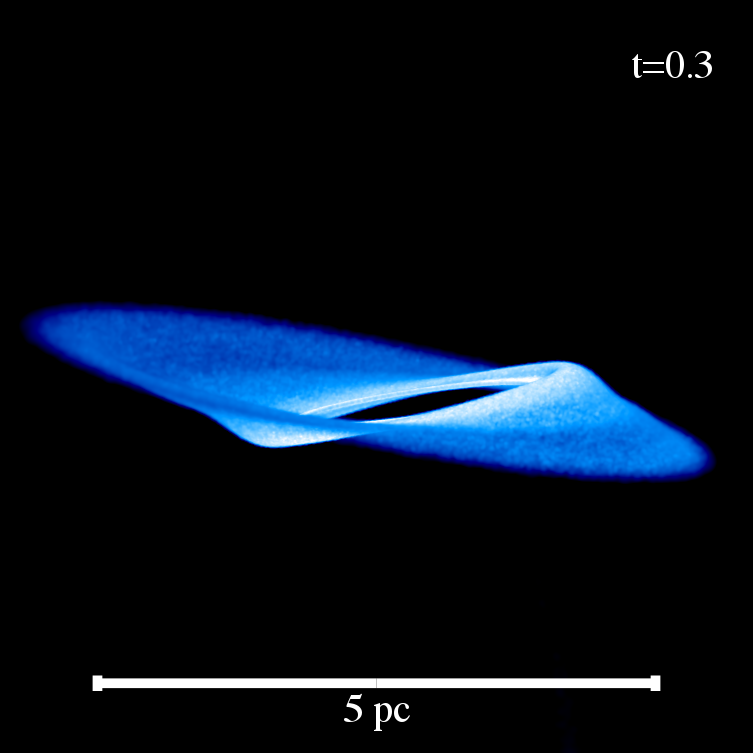}
\includegraphics[width=0.3\textwidth]{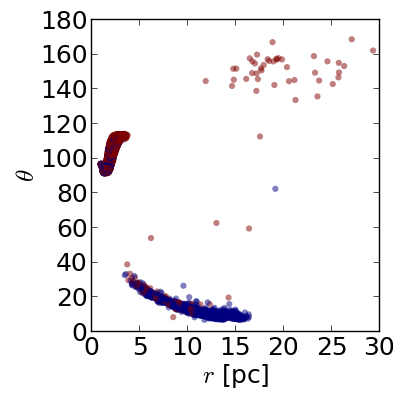}\\

\includegraphics[width=0.3\textwidth]{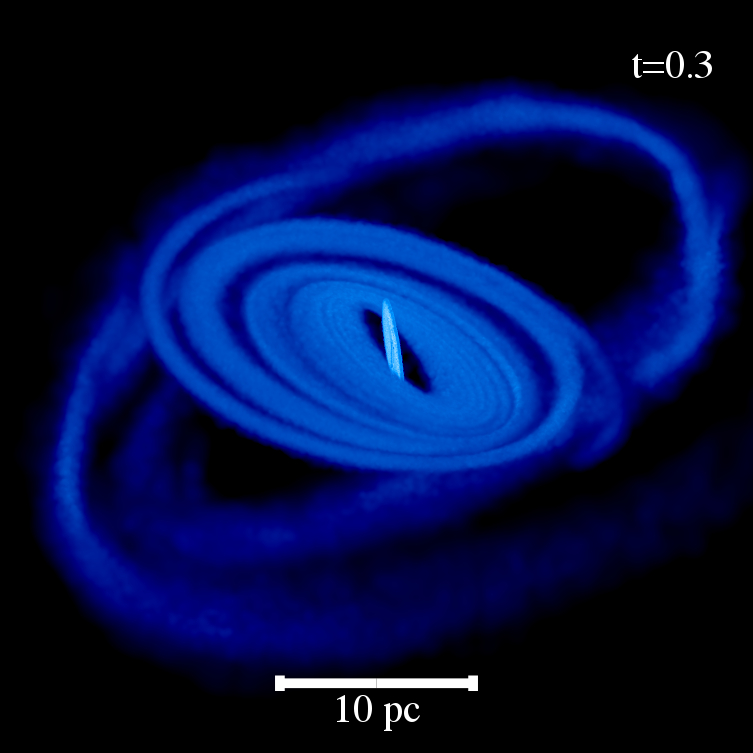}
\includegraphics[width=0.3\textwidth]{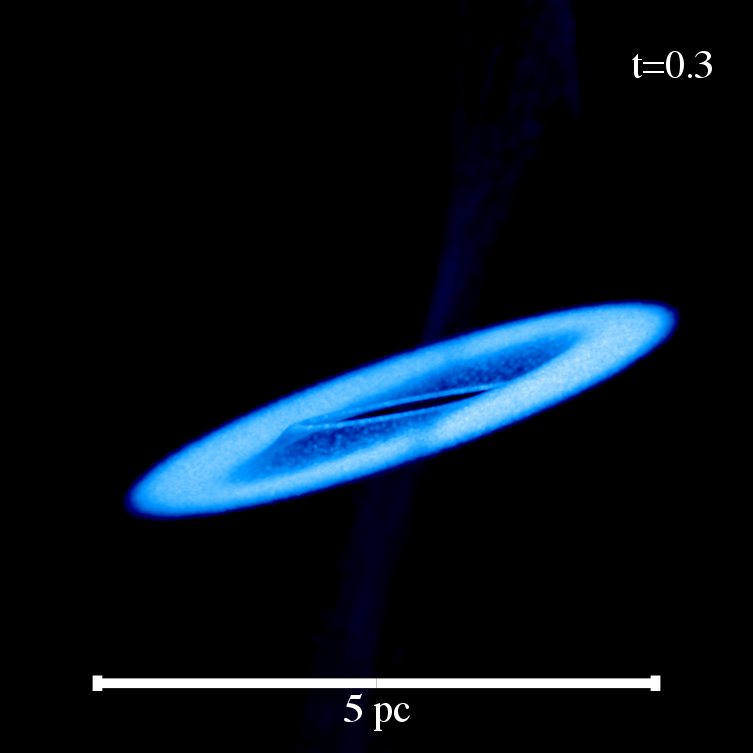}
\includegraphics[width=0.3\textwidth]{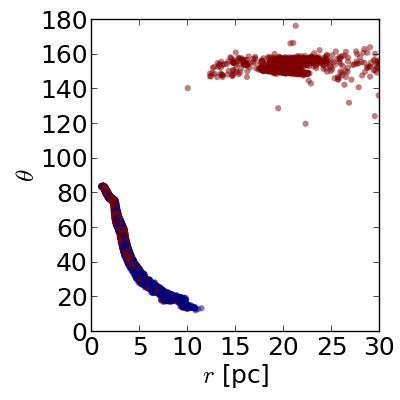}\\

\includegraphics[width=0.3\textwidth]{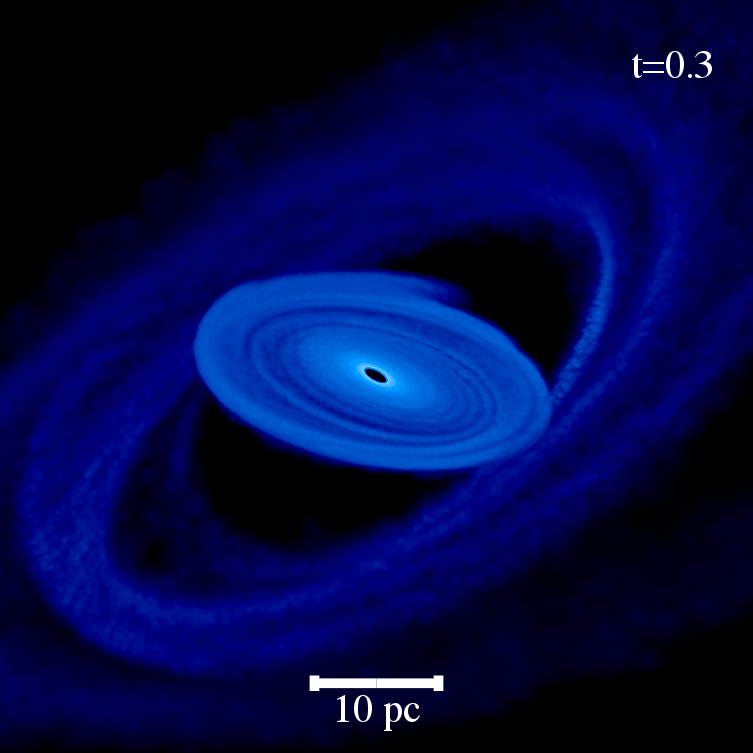}
\includegraphics[width=0.3\textwidth]{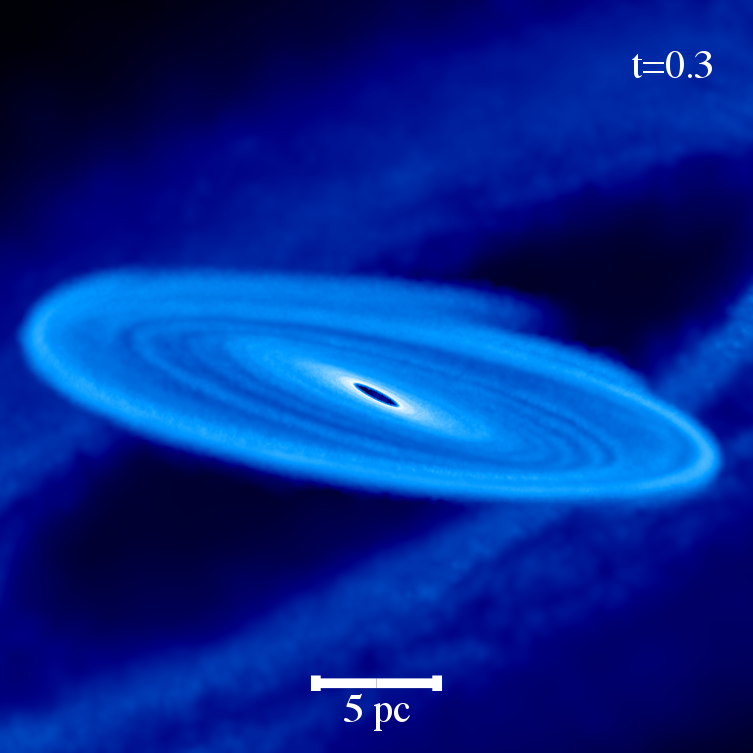}
\includegraphics[width=0.3\textwidth]{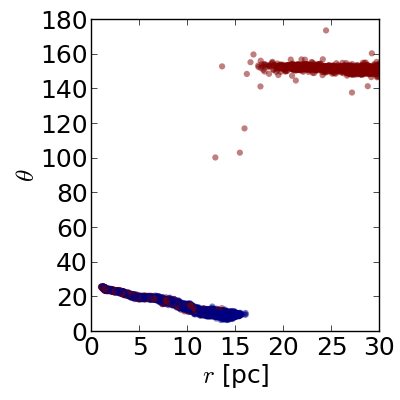}
\end{tabular}
\caption{Warped discs formed when an inflow of gas, here modelled as a rotating gaseous turbulent shell, impacts on a stalled disc in near \textbf{counter-rotation ($\tilt =150\degree$)}. The \textbf{left column} shows the generic outcome of these events. The \textbf{central column} shows a zoom into the innermost warped disc that forms (note that the disc view has been changed here to highlight better the degree of disc warping). The {\bf right column} shows 10\% of the particles selected randomly.  For each particle we record its distance  $r$ (in units of pc) from the SMBH and its tilt angle $\theta$ between the specific angular momentum of the particle and the $z$ axis (representing the rotation axis of the primitive disc). \textbf{Red particles} come from the shell while \textbf{blue particles} come from the primitive disc. We see that the inner and outer disc are always in counter-rotation. From top to bottom, the rows correspond to the cases v02t150, v03t150 and v07t150. The Figure illustrates the high degree of warping  of the structure.}
\label{fig: warps1}
\end{figure*}

Figure \ref{fig: warps1} shows the complex structure that forms after the hydrodynamical interaction, in the counter-rotating interaction cases\footnote{The snapshots of our simulated discs were done using \textsc{splash} (\citealt{PriceSplash07} )}. As the gas of the shell {\it rains} onto the disc, it shock-mixes with the fluid particles of the disc. The almost head-on encounter leads to an important cancellation of angular momentum, dragging a significant amount of mass within the accretion radius. Fluid particles that still preserve enough angular momentum form an extended,  nested {\it warped} disc comprising both shell and disc particles. In none of these cases do particles inside 5 pc rotate at the initial tilt angle ($\tilt = 0$). The right panels of Figure \ref{fig: warps1} show the angle $\theta$ as a function of spherical radius $r$ for comparing the strength of the warps.

\noindent
$\bullet$ In the run v02t150, the inflowing shell has a smaller angular momentum than that of the disc ($J_{\rm shell} = 2/3 \, J_{\rm disc}$).  Two decoupled warped discs are formed. One inside the 5 central parsecs, and the other extending from 5 to 15 parsecs approximately.  Both have lost memory of the initial angular momentum distribution. 

\noindent
$\bullet$ In run v03t150, the inflowing shell has the same angular momentum as the primitive disc ($J_{\rm shell} =  J_{\rm disc}$). At first sight (Figure \ref{fig: warps1}; left central panel), it seems that there are three disconnected discs. However, in fact only two decoupled discs are present (right central panel of Figure \ref{fig: warps1}): one warped disc extending from 10 pc inwards and an outer non-warped lighter disc extending from 15 to 30 pc. One would naively expect to have all the gas of the shell mixed with all the gas of the disc generating a single massive disc, shrunk and with its angular momentum making an angle $\theta \sim 1/2 \times 150\degree$ with the $z$ axis. However, since the interaction starts taking place as soon as the first gas from the shell impacts on the disc, the formation of the inner warped disc occurs before the rest of the shell has time to make its way down to the stalled disc. This late fraction of the shell still interacts with the rest of the disc, changing its orientation and shrinking its orbits but without producing significant inflow. At this time, because of the shrinkage of the primitive disc, particles of the shell that were on the high tail of the angular momentum distribution will not be able to interact any more, thus remaining in orbit as an outermost less dense disc with the same orientation the shell had initially ($\tilt = 150\degree$). 
  
\noindent
$\bullet$ In run v07t150, the inflowing shell has a higher angular momentum than that of the stalled disc ($J_{\rm shell} = 7/3 \, J_{\rm disc}$).  Two nested decoupled discs are formed with only the inner one being warped. The inner disc extends from the accretion radius out to 15 pc, while the outer non-warped disc extends from 15pc outwards (Lower panels of Figure \ref{fig: warps1}). The large kinetic mismatch between the two flows leaves behind a decoupled disc, with a  non trivial distribution of angular momentum which might reflect also how we set our initial conditions.

\subsubsection{Co-rotating inflows}

\begin{figure*}
\begin{tabular}{cc}
\includegraphics[width=0.3\textwidth]{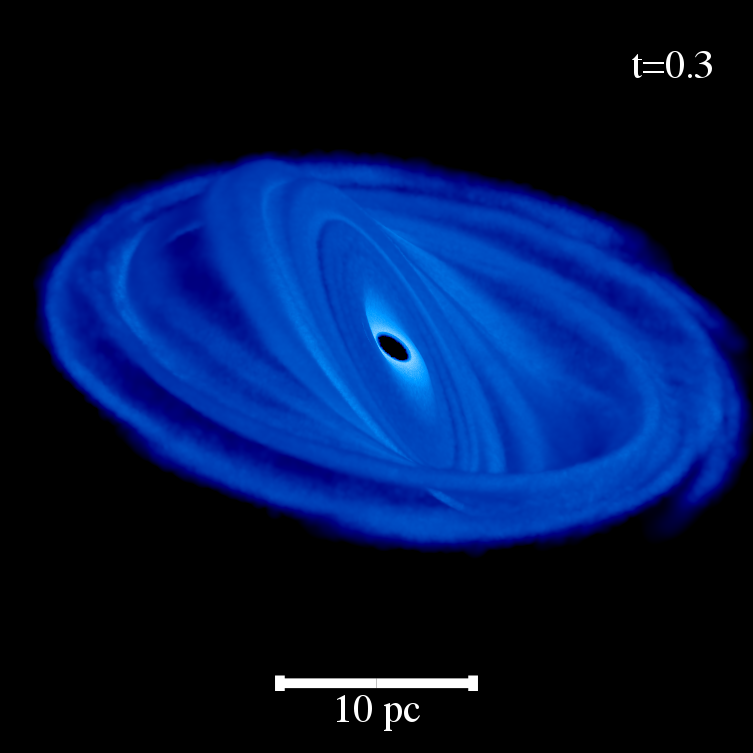}
\includegraphics[width=0.3\textwidth]{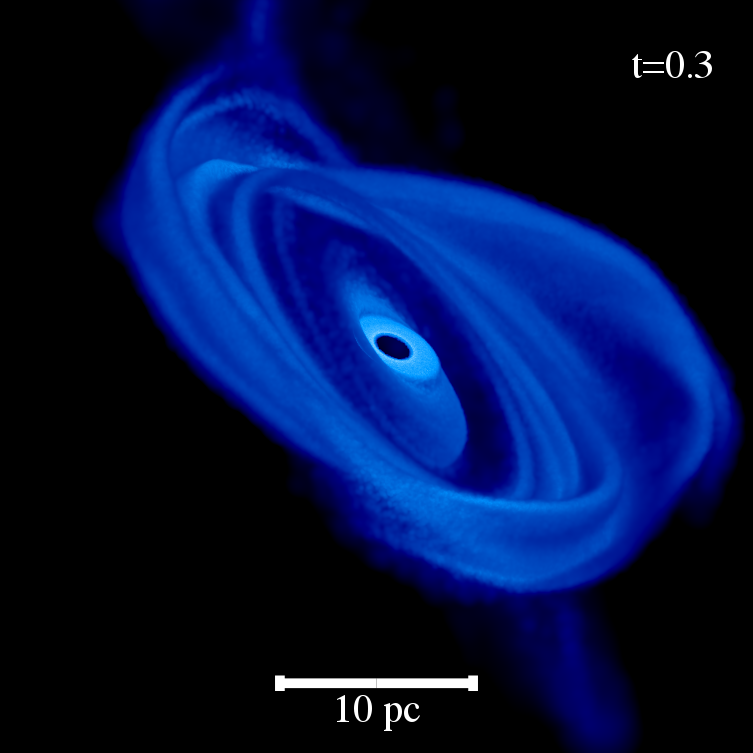}
\includegraphics[width=0.3\textwidth]{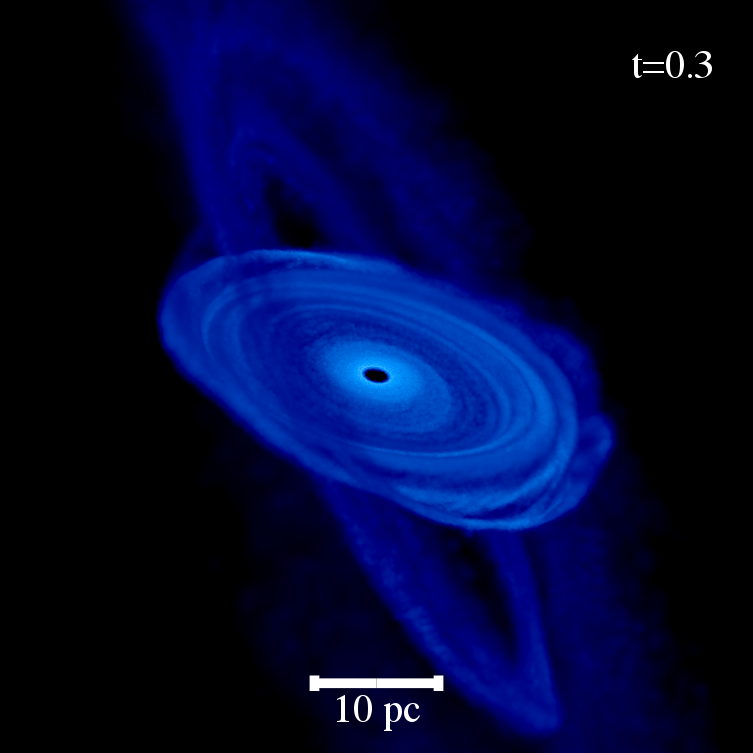}\\

\includegraphics[width=0.3\textwidth]{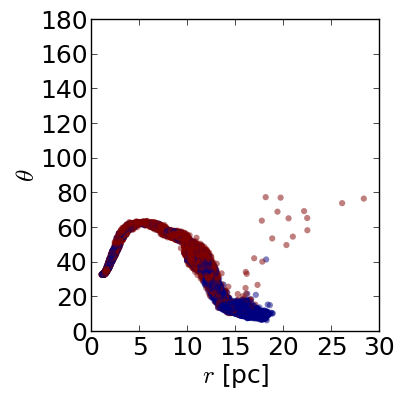}
\includegraphics[width=0.3\textwidth]{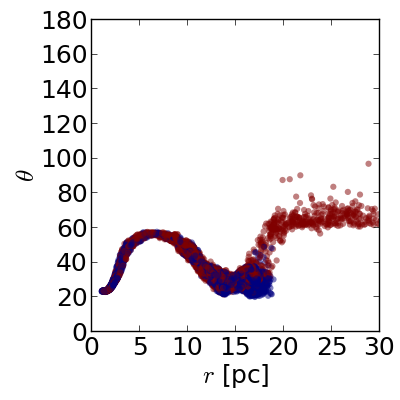}
\includegraphics[width=0.3\textwidth]{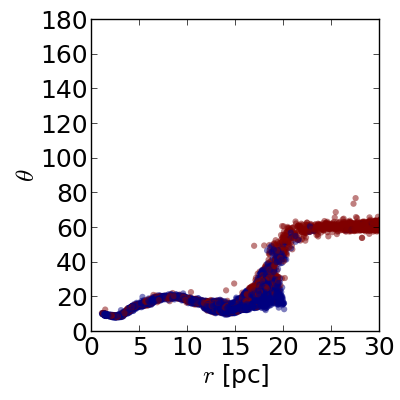}
\end{tabular}
\caption{Warped discs formed when an inflow of gas, here modelled as a rotating gaseous turbulent shell, impacts a primitive, stalled disc in near \textbf{co-rotation ($\tilt = 60\degree$)}. The {\bf upper row} shows the generic outcome of these events: a large-scale warped disc which changes its plane of rotation from the inside to the outside. The {\bf lower row} shows 10\% of the particles selected randomly.  For each particle we record its distance  $r$ (in units of pc) from the SMBH and its tilt angle $\theta$ between the specific angular momentum of the particle and the $z$ axis (representing the rotation axis of the primitive disc). \textbf{Red particles} come from the shell while \textbf{blue particles} come from the primitive, stalled disc. Note that these structures are always continuous discs changing their plane of rotation by as much as $\sim 60\degree$. From left to right, the columns correspond to the cases v02t060, v03t060 and v07t060. Note that for the case v07t060 it would seem that the inner part of the disc is detached from the outer part, however the $r-\theta$ plot shows otherwise. In all cases there is a continuous mix of shell and primitive disc particles.}
\label{fig: warps2}
\end{figure*}

When the infalling shell is nearly in near co-rotation with the stalled disc, there is no significant cancellation of angular momentum. The interaction leads to the formation of a single warped disc-like  structure in which $\theta$ changes orientation smoothly with radius, i.e. a larger scale warped disc forms as illustrated in  Figure \ref{fig: warps2}. For the cases in which the inflow has lower and higher angular momentum than the stalled disc (i.e. runs v02t060 and v07t060), the warp looks particularly clean, whereas for the cases in which the inflow and the primitive disc have the same angular momentum (run v03t060) the disc is more disturbed and the warp more structured. Note that $\theta$ is not changing monotonically with distance $r$ for any of the cases plotted.

For  run v03t060, a large number of particles is concentrated just below and just above the circularization radius ($\rcirc = 16\pc$) with rotation axes spanning a wide range of angles \textit{at the same radius} (see bottom middle panel of Figure \ref{fig: warps2}). These are particles whose rotation axis is either tilted by $\theta \simeq 10\degree$, almost parallel with the $z$ axis of the primitive disc, or tilted by $\theta \sim 50\degree$, almost aligned with the mean angular momentum of the inflow. This suggests that the interaction has not yet finished, even if most of the accretion has already taken place. Gas at these radii will continue to shock-mix until a state of equilibrium is reached in which the particles at each radius rotate around the same axis. 

Runs v02t060 and v07t060 are smoother. At each radius the particles rotate around a single axis (see Figure \ref{fig: warps2}, left and right lower panels). From the inside to the outside the rotation axes in both cases span almost $\Delta \theta \simeq 60 \degree$.

As a general trend, gas at the circularization radius of the primitive, stalled disc has its angular momentum aligned with the $z$ axis while gas at the circularization radius of the shell has its angular momentum tilted with respect to the $z$ axis by $\tilt$, keeping memory of its original orientation.

\section{Observational signatures}
\label{sec: OS}

\begin{figure}
\includegraphics[width=0.23\textwidth]{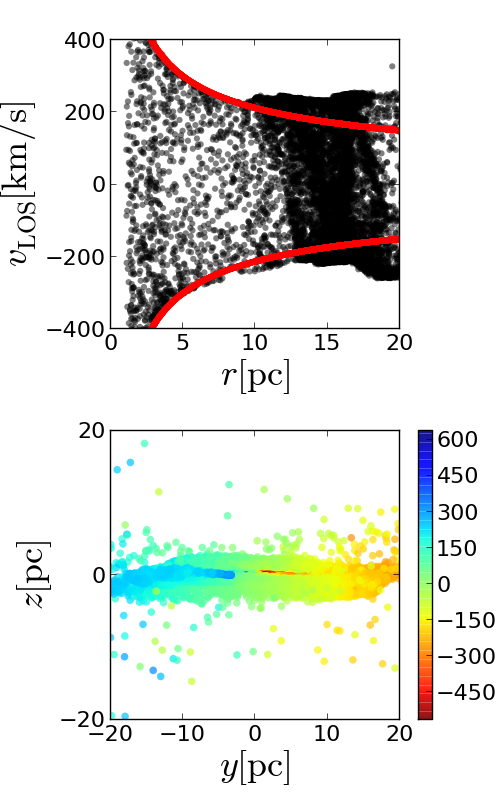}
\includegraphics[width=0.23\textwidth]{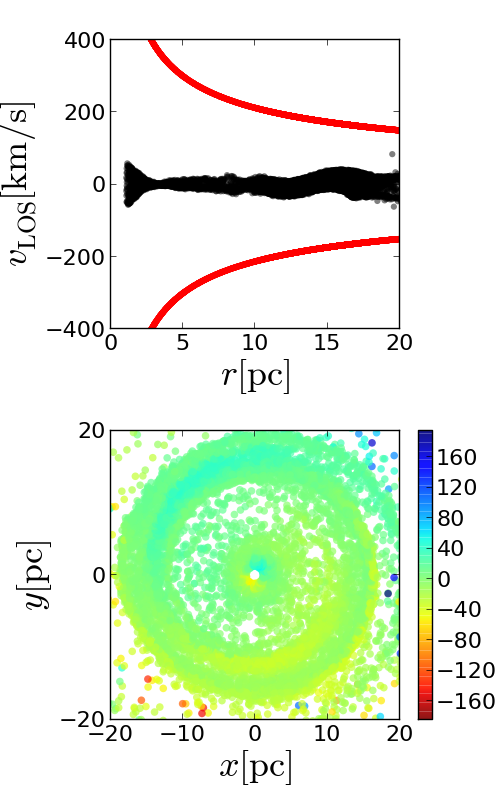}
\caption{Single disc viewed edge-on (left) and face-on (right). In the top panels the line of sight (LOS) velocity of each particle is plotted in black against its distance from the SMBH. Keplerian LOS velocity profiles are superimposed in red. Note that in both cases the velocity profiles of the particles are consistent with Keplerian rotation: $v_{\rm LOS} \propto r^{-1/2} \cos(i)$, where $i = 0$ when the disc is edge-on and $i = 90$ when it is face-on. In the lower panels, we see the LOS velocity map. Bluer colours represent gas approaching the observer while redder colours represent gas which is receding. In the edge-on case, the LOS velocity goes from being positive on the left side to negative on the right side (with an absolute maximum value of $\sim 500\, \kms$), while in the face-on case the LOS velocity is very small.}
\label{fig: LOSvel1}
\end{figure}

\begin{figure}
\includegraphics[width=0.23\textwidth]{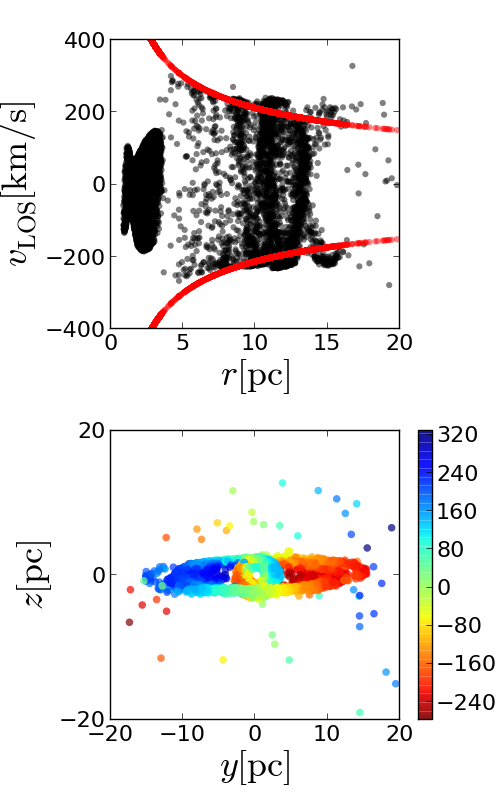}
\includegraphics[width=0.23\textwidth]{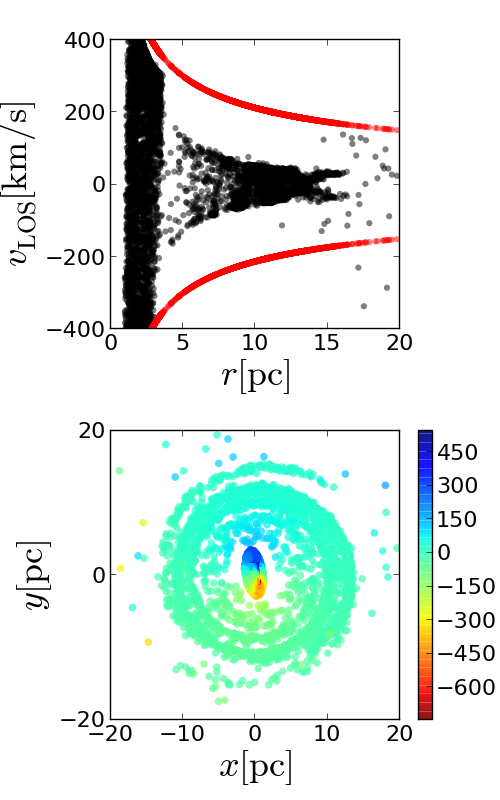}
\caption{Warped discs that form after the shell-disc interaction for the case vrt02tilt150 (the shell which fell with lower angular momentum than the disc and in near counter-rotation). The lines of sight are the same as  in Figure \ref{fig: LOSvel1}. In the top panels the line of sight (LOS) velocity of each particle is plotted in black against its distance from the SMBH. Keplerian LOS velocity profiles are superimposed in red. Note that in both cases the velocity profiles of the particles are not consistent with a flat disc in Keplerian rotation: $v_{\rm LOS} \propto r^{-1/2} \cos(i)$, where $i = 0$ when the disc is edge-on and $i = 90\degree$ when it is face-on. In the lower panels, we see the LOS velocity map. Bluer colours represent gas approaching the observer while redder colours represent gas receding from her/him. In both of the bottom panels it is clear that in the central region the gas is kinematically detached from the rest.}
\label{fig: LOSvel2}
\end{figure}

A comparison between the rotation profiles of a single disc and of a warped disc formed after the interaction of two nested inflows is carried out in  Figures \ref{fig: LOSvel1} and \ref{fig: LOSvel2}, where we show  a random sample of 10 \% of the particles for two selected simulations: a single disc generated by an infalling turbulent gaseous shell (run v03) and the end state of a shell that fell onto a stalled disc with lower angular momentum and in near counter-rotation (run vrt02t150). The left and right panels in each figure correspond to the gas observed from two perpendicular lines of sight (LOS) that coincide with the primitive disc being edge-on and face-on respectively. In the top panels the LOS velocity of each particle is plotted against its radial distance. In addition, a red line showing the velocity profile, $v_{\rm K} \propto r^{-1/2} \cos(i)$, expected for a Keplerian disc, is superimposed,  where the angle $i = 0$ for an edge-on disc and $i = 90\degree$ for a face-on disc. From these, we immediately note that the single turbulent disc (Figure \ref{fig: LOSvel1}) has a rotation profile consistent with an almost flat Keplerian disc, whereas the composite case of a shell in counter-rotation impacting on the disc (Figure \ref{fig: LOSvel2}) shows a very different profile, inconsistent with Keplerian rotation. In the lower panels, LOS velocity maps are shown with red and blue colours indicating receding and approaching gas. In the single disc case (Figure \ref{fig: LOSvel1}), a very ordered rotation is observed: gas approaching from the left and receding to the right when viewing the disc edge-on, and gas with a very tiny almost constant velocity when the disc is face-on. Instead, when observing the remnant of the overlap (Figure \ref{fig: LOSvel2}), the ordered pattern just described is not present and it is clear that two kinematically decoupled bodies of gas are present. 

\textbf{The comparisons just made must be taken only as an indication of the kind of signatures expected. To obtain more realistic signatures, one should take into account a live stellar potential which would certainly exert torques on the gas, thus contributing to the final shape of the warp. An additional interesting problem is that of exploring the possible formation of \textit{stellar} warped discs, which must be tackled including in-situ star formation.}

\section{Discussion}
\label{sec: discussion}

Previous works have examined the collision of a gas cloud with a pre-existing nuclear gas disc around a black hole (\citealt{Alig13}), the collision of two gas clouds orbiting a central black hole (\citealt{Hobbs09}) and the impact of a prolate gas cloud impinging and engulfing a central black hole (\citealt{Lucas13}) by means of idealised hydrodynamical simulations with the particular aim of explaining the origin of the disc of young stars observed at the Galactic Centre around Sgr. A* (e.g. \citealt{Genzel10}). The common feature of all of these experiments is the collision of gas flows with different angular momenta and the subsequent formation of warped and/or nested discs. If the Galactic Centre represents a standard, these observations complemented by theoretical modelling suggest that galactic nuclei are sites where collisions between gas clouds and residual discs take place. The model which we have presented here and in Paper-I can be regarded as related to these processes. It has a higher degree of symmetry than those but also generates warped and nested rings. It seems plausible then, that a relation between the mass inflow and the angular momentum difference between two gaseous components (disc-cloud, cloud-cloud) can be obtained, even if in an approximate way. This could be relevant for more accurate sub-grid models of AGN feedback in cosmological simulations (e.g. \citealt{Gaspari13}).

Strong circumstantial evidence indicates that warps are common in AGN accretion discs. For instance, maser emission in the nuclear regions of NGC 4258 (\citealt{Herrnstein05}), Circinus (\citealt{Greenhill03}), and four of the seven galaxies examined by \cite{Kuo11} has revealed the presence of warped gaseous discs at 0.1-1 pc scales. Additionally, warped thin discs have been invoked by \cite{Nayakshin05} as the obscuring mechanism behind the unification paradigm of AGN. They argue that warped clumpy discs provide the easiest way to obscure the central engine of AGN. 

All of this suggests that a connection between warped discs and AGN fuelling is not just accidental. \textbf{However, we need to keep in mind that, in order to establish a connection between the warping of the disc and AGN fuelling, a realistic treatment of the stellar potential must be included, as gravitational torques exerted by it could potentially modify the structure of the warps and the AGN fuelling rate.}

\section{Conclusions}
\label{sec: concl}

The presence of a stalled gas disc around an SMBH on scales between $\sim1-30 \pc$, where the angular momentum barrier is found, can enhance the inflow of matter that subsequently reaches these regions. By interacting via shocks with the stalled disc, angular momentum cancellation takes place. In this way, the stalled disc acts as an agent or ``obstacle'' that modifies the free flow of any gas reaching it. Misalignment between the angular momenta of the arriving gas and that of the disc is preferred for an efficient cancellation of angular momentum. Such a situation is represented in our suite of simulations by the shell reaching the disc in counter-rotation, which greatly enhances inflow (see Figure \ref{fig: compare}). An even greater inflow is obtained in cases in which the shell is initially given a turbulent velocity field.

The interactions described in this work generate nested and kinematically detached warped discs and establish a natural connection between the angular momentum cancellation mechanism and the warping of the discs. Co-rotating inflows give origin to smoother warps that are nonetheless interesting to study. Such warped discs on $\sim 10$ pc scales, show line-of-sight velocity profiles that depart significantly from Keplerian rotation, as illustrated in Figure \ref{fig: LOSvel2}. These warped structures, on parsec  scales, arise from the purely hydrodynamical interaction between a misaligned  inflow and a stalled gas disc. Neither SMBH feeding nor disc warping in this scenario invoke the action of gravitational or viscous torques (see e.g. \citealt{Tremaine14}). More realistic simulations of e.g. cloud-cloud collision, motivated by state of the art simulation of galaxies under formation and evolution will asses whether the ``cancellation of angular momentum by a stalled gaseous disc'' is an important catalyst of SMBH feeding and AGN fuelling. \textbf{At the same time, more realistic simulations including a live stellar potential will quantify the relative impact on the disc's shape of the overlapping inflows scenario with respect to the gravitational torques generated by a non-spherical stellar bulge.}


\acknowledgments
\noindent The authors wish to thank Klaus Dolag for making available the code \pgadget and John Miller for his valuable comments and critical reading of the manuscript. JMCL thanks Giuseppe Murante and Sandra Raimundo for enlightening discussions.

\vskip 24pt

\bibliographystyle{apj}
\bibliography{TurbulentOverlapping}


\appendix

\section{Self-Gravity}
\label{sec: SG}
In this Appendix we explore the role of self-gravity by varying the mass of the gaseous inflowing shell which will then settle into the ``stalled'' disc. For very low shell masses as compared to the mass of the SMBH, $m = \mshell/\mbh << 1$, we expect that the inflow will proceed cleanly as if the fluid were non self-gravitating. With increasing $m$, we expect the formation of clumps due to local gravitational instabilities. By doing this exercise, we determine the mass threshold below which self-gravity  is negligible, and at the same time we assess the role of self-gravity in altering the values of the inflow rate onto the SMBH in our simulations. 

In a massive gaseous disc, the parameter that tilts the balance towards fragmentation or stability is the cooling time. Indeed, for a disc to be stable against self-gravity, the \cite{Toomre64} parameter,

\begin{equation}
Q = \frac{c_s\,\kappa}{\pi\,G\,\Sigma},
\end{equation}

\noindent where $\kappa$ and $\Sigma$ are the epicyclic frequency and the surface density respectively, must be greater than unity, with marginal stability occurring at $Q = 1$.  When the cooling time-scale of the gas, $t_{\rm cool}$, is greater than the dynamical time-scale of the disc, $t_{\rm dyn}$, an initially hot disc ($Q>1$) will cool slowly until the marginal instability stage is reached ($Q=1$). During this process, the disc has had enough time to respond, with local gravitational perturbations forming into a global spiral structure that in turn heats the gas through shocks and $p\,dV$ work, preventing the disc from going into the fragmenting regime ($Q < 1$). Thus, the interplay between local self-gravity (governed by $t_{\rm dyn}$) and the cooling of the gas (governed by $t_{\rm cool}$) will dictate the evolution of the disc. In the scenario just described ($t_{\rm cool} > t_{\rm dyn}$), the spiral structure acts as a thermostat, always maintaining the gas at marginal stability. On the other hand, if $t_{\rm dyn} > t_{\rm cool}$, as the disc cools faster than the time which it needs to respond and develop a spiral structure, the disc breaks into clumps (for a more detailed discussion of these processes see e.g. \citealt{Lodato07, Lodato12}). 

\begin{figure}
\includegraphics[width=0.49\textwidth]{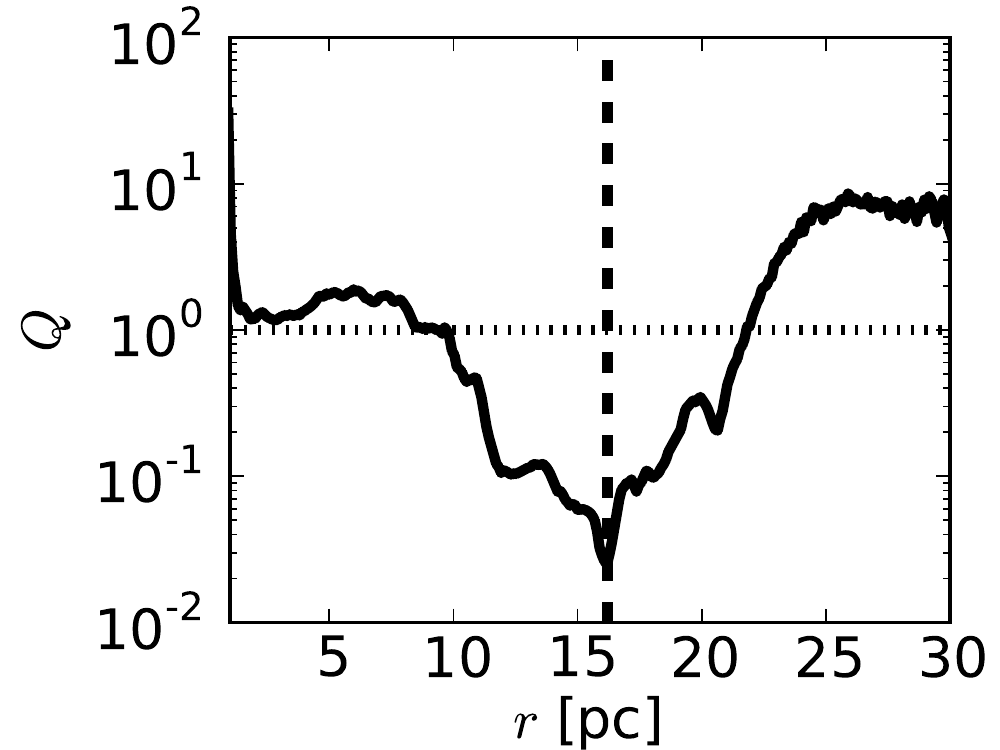}
\caption{The Toomre parameter calculated for a non self-gravitating thin disc in Keplerian rotation around the SMBH of mass $\mdisc = 0.1 \mbh$. In this case, the surface density $\Sigma(r) \sim r^{-3/2}$, the density $\rho(r) \sim r^{-3}$ and the total scale height $H(r) = 0.02 \, r$. With this numbers at hand, the Toomre parameter can be estimated as $Q \simeq \mbh / [\pi \, r^3 \, \rho(r) ] \simeq k \, \mbh/\mdisc$, with $k \simeq 0.1$. The Toomre parameter is roughly constant within the inner 10 parsecs and reaches a minimum $Q << 1$ at the circularization radius $\rcirc \simeq 16 \pc$. Note that the same shape $Q(r)$ shifts upwards or downwards depending on $\mdisc$.}
\label{fig: Toomre}
\end{figure}

Cold constant-temperature discs, having an idealized $t_{\rm cool} = 0$, are thus prone to fragmentation. In this way, the discs in our simulations, with a temperature $T = 10^3 K$, stand on the worst-case scenario in which there is no way of halting fragmentation once the disc becomes unstable. In a constant-temperature thin disc with a Keplerian rotational profile ($\kappa \simeq \Omega(r)$), $Q$ depends on the mass of the disc only through its surface density, $\Sigma$. If we were to include a more realistic cooling treatment, more massive discs would still be able to survive fragmentation, making it easier to feed the black hole through overlapping inflows and thus strengthening our conclusions. The Toomre parameter calculated for a non self-gravitating disc is shown in Figure \ref{fig: Toomre}. The minimum of the Toomre parameter, $Q_{\rm min}$, occurs at the circularization radius ($\rcirc \simeq 16 \pc$) due to the greater concentration of mass there.

We performed 4 simulations of single turbulent inflows in which the total mass in the gaseous shell was varied, taking into consideration the self-gravity of the shell. The initial conditions were the same as those described in section \ref{sec: IC}, in these specific cases with an initial rotation velocity of $\vrot = \vturb = 0.3$ and a mass $M_{\rm shell} = s \times M_{\rm BH}$, for $s = 0.01, \,0.05, \,0.1, \,1.0$.

\vspace{-12pt}
\subsection{The development of clumps following the infall of a single shell}

\begin{figure*}
\begin{tabular}{cc}
\includegraphics[width=0.49\textwidth]{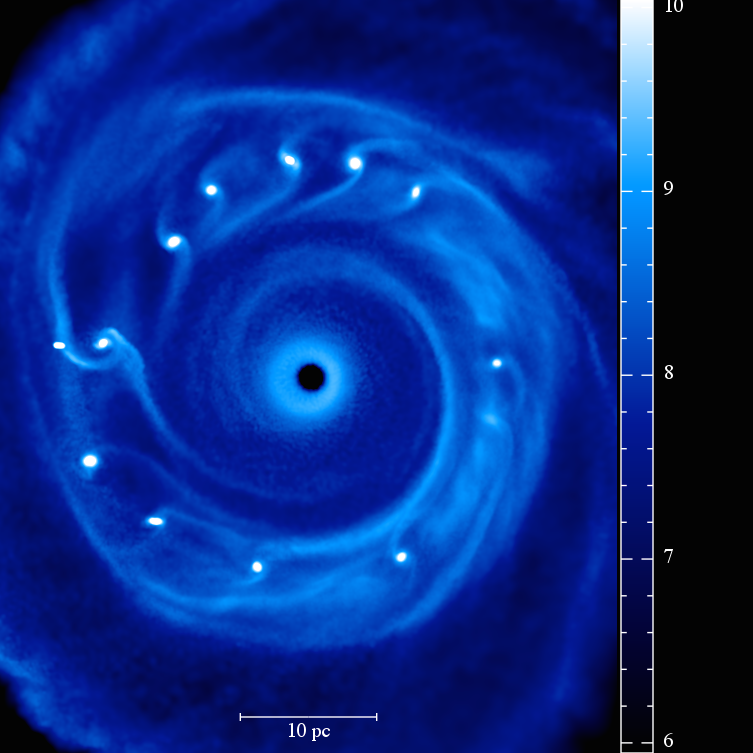}\,
\includegraphics[width=0.49\textwidth]{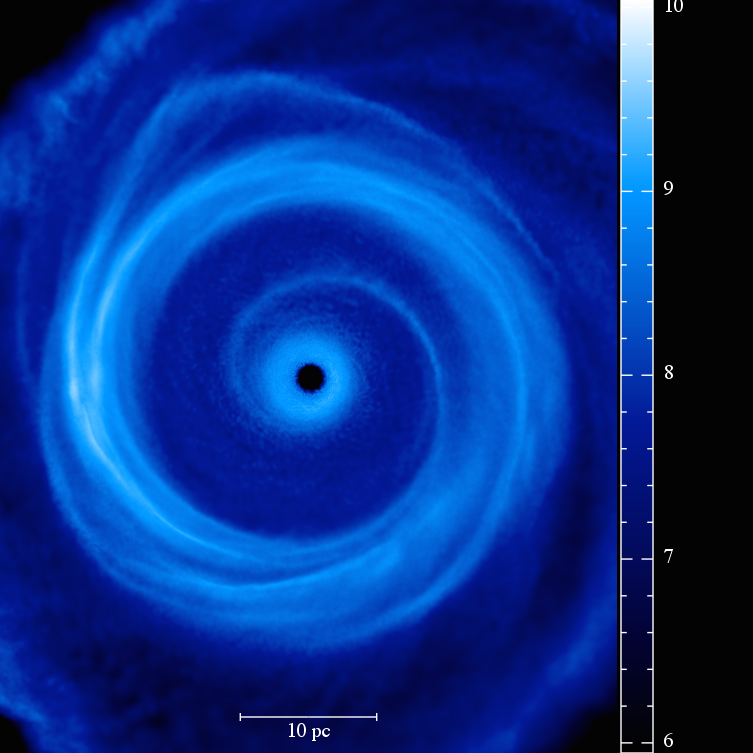}
\end{tabular}
\caption{Projected $\log (\Sigma/\mshell)$ in the $xy$ plane when a disc form ($t =0.3$ expressed in time units $T_u=5$ Mry). The left-panel shows self-bound clumps in the
 $\mshell = 0.1\,\mbh$ model where self-gravity is important. The right-panel shows the disc in model 
$\mshell = 0.01\,\mbh$ where self-gravity is unimportant.}
\label{fig: clumpslater}
\end{figure*}

Clumps developed in all of the simulations of this section with the exception of the lowest mass run $m=0.01.$
In the left-panel of Figure \ref{fig: clumpslater}  we show clump formation for the case $m=0.1$  at the time of disc formation ($t=0.3$).
The clumps develop within a narrow ring at a circularization radius, $r_{\rm circ} \simeq 16 \pc$ (as could be predicted by Figure \ref{fig: Toomre}). The clumps become gravitationally bound and revolve around the SMBH in quasi-Keplerian orbits. As they form at the circularization radius due to the concentration of mass at that radius, their rotation around the SMBH is quite stable, so that they barely interact with each other. When they do interact, they merge into  larger and denser clumps, but no scattering occurs that can bring mass towards the SMBH. Tidal tails develop among them but these are not strong enough to disrupt the dynamics of the disc outside their vicinities.
In the right-panel of Figure \ref{fig: clumpslater}  we show the smooth disc that forms when 
$m=0.01$ corresponding to the case of negligible self-gravity.

\vspace{-12pt}
\subsection{Inflow Rates}

\begin{figure}
\includegraphics[width=0.49\textwidth]{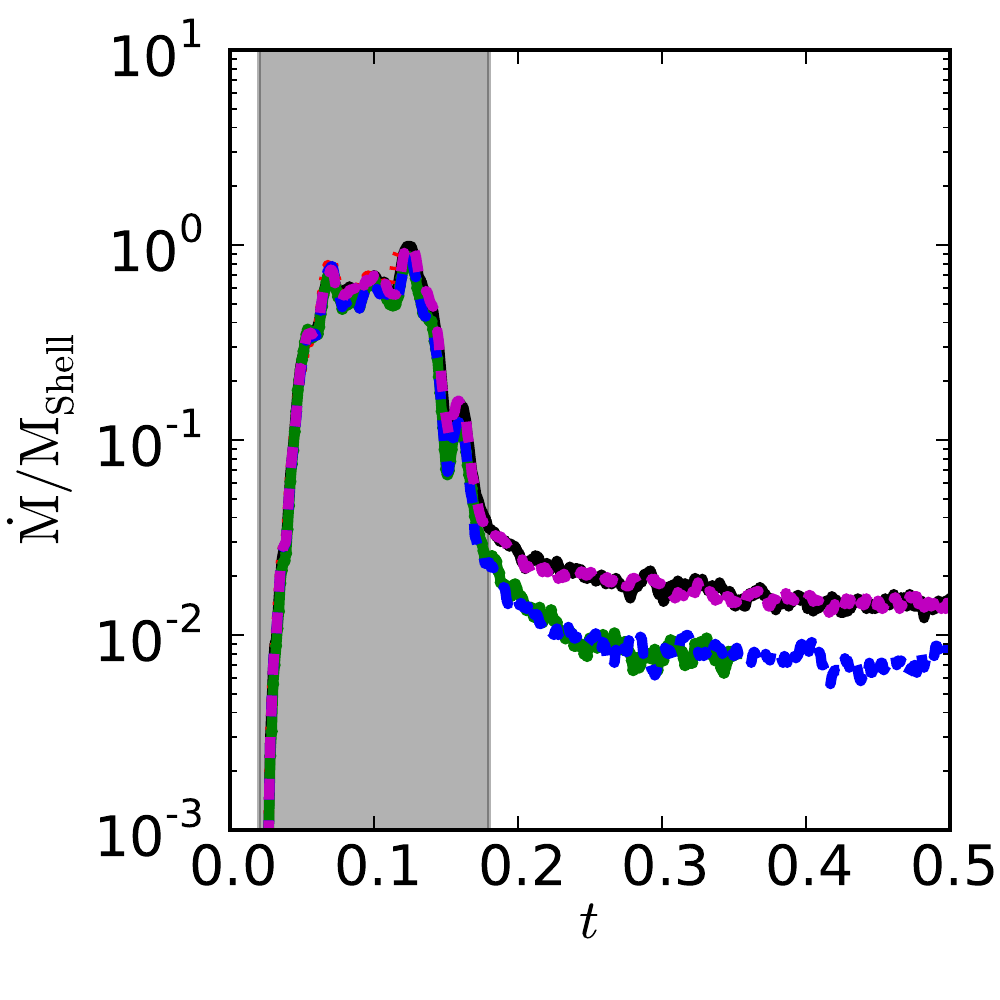}
\caption{Particle accretion rates as a function of dimensionless time for the cases in which the mass of the inflowing shell is $M_{\rm shell} = \mbh$ (red dotted), $0.1 \mbh$ (blue-dashed), $\mshell = 0.05 \mbh$ (green dashed)  and $\mshell = 0.01 \mbh$ (magenta-dashed). For comparison, the particle accretion rate for a case in which the non self-gravity approximation was used is also shown (black solid line). The overlap of the lines makes it nearly impossible to distinguish between them at relevant levels of accretion ($t \lsim 0.18$; shaded region), which correctly illustrates that particle accretion rates are independent of the mass of the shell.}
\label{fig: v03SGaccrate}
\end{figure}

We now compare the specific inflow rates, $\dot{M}/\mshell$, toward the central parsec, produced by varying the mass of the inflow. By doing this we can have a grasp of how relevant self-gravity is in either helping or preventing gas from falling from scales of $10 -100 \pc$ down to the central parsec. The specific inflow rates for the four cases are compared in Figure \ref{fig: v03SGaccrate}. It appears that the specific accretion rates are independent of the mass of the inflow when the masses are low enough, $M_{\rm shell} < 0.05 \mbh$. Even in the cases with $\mshell = 0.05-0.1\,\mbh$, in which self-gravity starts playing an important dynamical role by disrupting the inflow into clumps at $\rcirc$, the accretion rate during the most active phase, $0.03 \lsim t \lsim 0.2$, remains unaffected by self-gravity. This means that we can safely model inflows of single shells as being non-self-gravitating whenever $\mshell \lsim 0.1 \mbh$, as far as the sub-pc inflow rate is concerned. This threshold must change when dealing with overlapping inflows, as we must have a  space-filling gaseous disc for the second inflow to interact with. Therefore, when treating overlapping inflows, we must have $M_{\rm tot} = \mdisc + \mshell \lsim 0.05 \mbh$.
 
\end{document}